\documentclass{sigchi-ext}
\usepackage[T1]{fontenc}
\usepackage{textcomp}
\usepackage[scaled=.92]{helvet} 
\usepackage{graphicx} 
\usepackage{balance}  
\usepackage{booktabs} 
\usepackage{ccicons}  
\usepackage{ragged2e} 

\def\plaintitle{Challenges in Supporting Exploratory Search through Voice Assistants}

\def\emptyauthor{}
\def\plainkeywords{Voice user interface; conversational user interface; voice-based search; voice assistants; digital assistant; virtual assistant; informational search; exploratory search}

\title{\plaintitle}

\numberofauthors{2}
\author{%
  \alignauthor{%
    \textbf{Xiao Ma}\\
    \affaddr{Google} \\
    \affaddr{New York, NY 10011, USA} \\
    \email{xmaa@google.com} 
  }
  \alignauthor{%
    \textbf{Ariel Liu}\\
    \affaddr{Google} \\
    \affaddr{Mountain View, CA 94043, USA} \\
    \email{arielliu@google.com} 
  }
}

\definecolor{linkColor}{RGB}{6,125,233}
\hypersetup{%
  pdftitle={\plaintitle},
  pdfauthor={\emptyauthor},
  pdfkeywords={\plainkeywords},
  bookmarksnumbered,
  pdfstartview={FitH},
  colorlinks,
  citecolor=black,
  filecolor=black,
  linkcolor=black,
  urlcolor=linkColor,
  breaklinks=true,
}

\begin{document}

\CopyrightYear{2020}
\setcopyright{rightsretained}
\conferenceinfo{CHI'20,}{April  25--30, 2020, Honolulu, HI, USA}
\isbn{978-1-4503-6819-3/20/04}
\doi{}
\copyrightinfo{\acmcopyright}

\maketitle

\RaggedRight{} 

\begin{abstract}

Voice assistants have been successfully adopted for simple, routine tasks, such as asking for the weather or setting an alarm.
However, as people get more familiar with voice assistants, they may increase their expectations for more complex tasks, such as exploratory search --- e.g., ``What should I do when I visit Paris with kids? Oh, and ideally not too expensive.''
Compared to simple search tasks such as ``How tall is the Eiffel Tower?'', which can be answered with a single-shot answer, the response to exploratory search is more nuanced, especially through voice-based assistants.
In this paper, we outline four challenges in designing voice assistants that can better support exploratory search: addressing situationally induced impairments; working with mixed-modal interactions; designing for diverse populations; and meeting users' expectations and gaining their trust.
Addressing these challenges is important for developing more ``intelligent'' voice-based personal assistants.

\end{abstract}





\section{Introduction}

The increasing adoption of voice assistants has been discussed in a considerable amount of academic literature~\cite{schalkwyk2010your,guy2018characteristics,ammari2019music,bhalla2018exploratory}.
We use the term ``voice assistants'' to refer to digital or virtual assistants whose interactions are primarily driven by speech, but ``voice assistants'' can also support other modes of interactions including visual and gestural.
In particular, prior work suggests that voice assistants are commonly used for music and for productivity tasks (setting alarms and timers), as well as in Internet of Things (IoT) devices and home automation~\cite{ammari2019music}.

\begin{figure}
  \centering
  \includegraphics[width=0.7\linewidth]{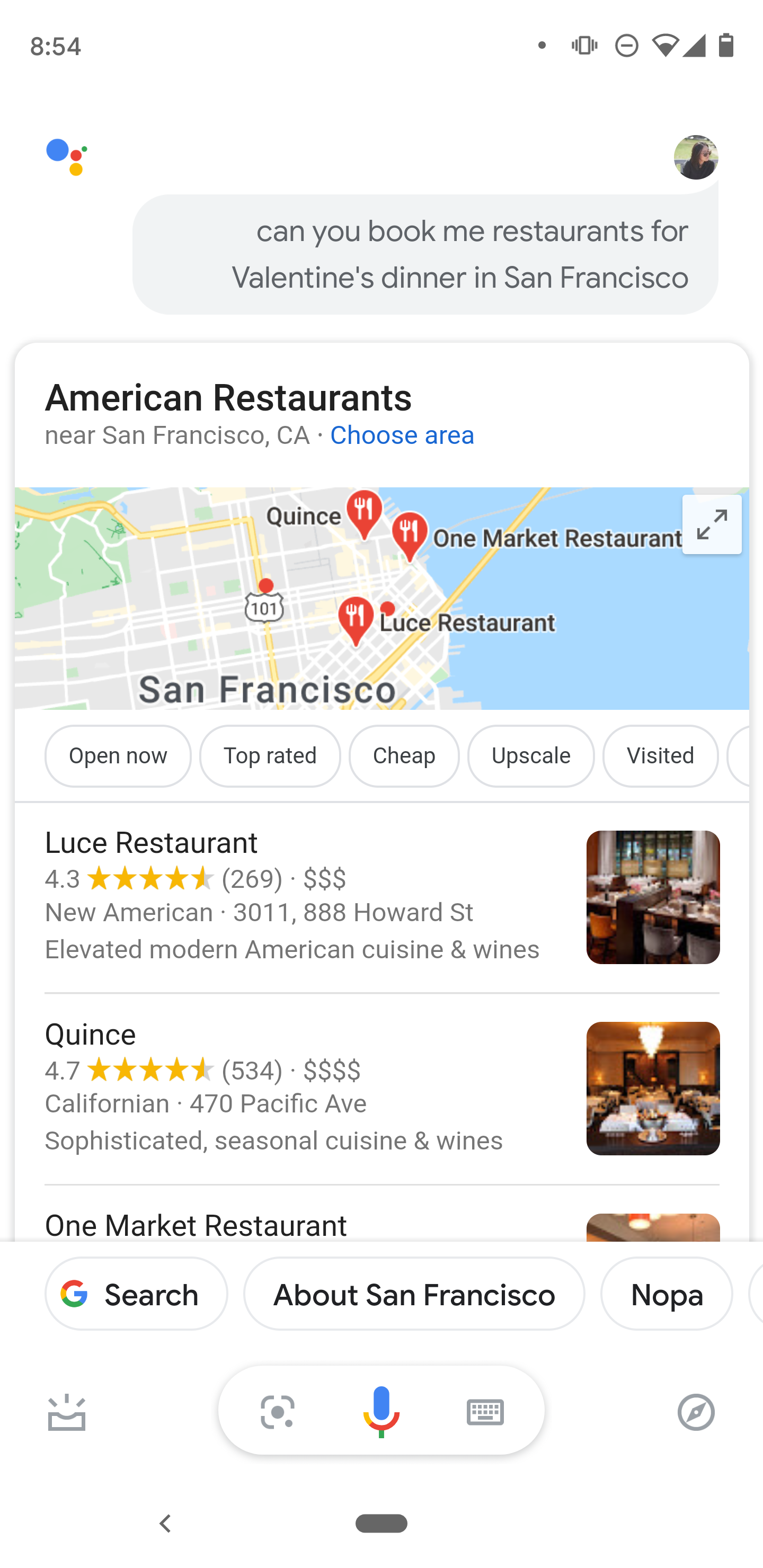}
  \caption{Voice assistant returns a single-shot response for exploratory search. A more natural dialog flow may ask follow-up questions to further help users understand and refine results.}
  \label{fig:exploratory_search}
\end{figure}

In the meantime, one increasingly important use case of voice assistants, and perhaps the holy grail of intelligent agents, is \emph{exploratory search}.
Exploratory search, in information retrieval literature, refers to search activities that are pertinent to ``learn'' and ``investigate', as opposed to ``look up''~\cite{marchionini2006exploratory}.
For example, a user asks the voice assistant for ideas for restaurants on Valentine's Day, with constraints on location, budget, availability, and ambience that haven't been thought-through yet~\cite{luger2016like}.
Current voice assistants treat such exploratory search similar to factual search (e.g., How tall is Obama?), returning a single-shot response of a list of retrieved results (see Figure \ref{fig:exploratory_search}).
However, in a more natural dialog flow, voice assistants can ask follow-up questions that help users learn, understand, and refine their choices.

There are many open questions and challenges in designing for supporting exploratory search through voice assistants.
In this position paper, we aim to lay out four of the challenges: addressing situationally induced impairments; working with mixed-modal interactions; designing for diverse populations; and meeting users' expectations and gaining their trust.
Addressing these challenges can help us collectively drive toward more capable voice assistants.

\section{Challenges in Voice Exploratory Search}

\subsection{Addressing Situationally Induced Impairments}
Voice assistants are inherently hands-free, voice-forward and of a mobile nature.
Hence, voice-based interactions are often used while dealing with different kinds of situationally induced impairments --- when it is not convenient to touch a phone or digital device, e.g., when walking, biking, driving, cooking, or eating~\cite{vtyurina2019towards}.
Situationally induced impairments create challenges for voice exploratory search due to high cognitive load for audio perception compared to visual perception~\cite{vtyurina2019towards}, but at the same time makes voice-based interactions even more valuable because there is no alternative.

The key tension in voice-based exploratory search during situationally induced impairments is balancing \emph{speed} and exploration.
In the example in Figure \ref{fig:exploratory_search}, each follow-up question around refinements (e.g., budget, availability, location, etc) together with user response to each question would add a significant amount of time to the interaction.
It is an open question to quantify exactly how much more time it would take to complete the same refinement flow using voice-based interactions, compared to touch- or click-based UI elements.
Such quantification can guide the design trade-offs when it comes to speed versus adding more support for exploration.


Another challenge for voice-based interaction as a result of situationally induced impairments is \emph{privacy}.
Voice search often take place in public environments, such as at a bus stop or on the street~\cite{cho2019hey}.
Compared to touch- or click-based interactions, voice makes users more vulnerable when the information being exchanged is personal or sensitive.
For example, if a user is searching for clothing and needs to explore the size, it might be particularly challenging to answer ``So what is your waist width?'' through voice than text.

\subsection{Working with Mixed-Modal Interactions}

Going beyond situationally induced impairments, when users can interact both through voice and visual-based elements, one of the core challenges that follow is deciding how to ``mix'' mixed-modal interactions.
Mixed-modal refers to the fact that although voice assistants are voice-forward, the responses they return could be a mix of speech, text, and visual elements.
At the same time, user input can also be through a mix of speaking, typing, clicking, and even more.
Compared to using voice alone, mixed-model interaction could be an effective design paradigm to improve information throughput leveraging visual elements and interactive UI.
For example, when a user asks, ``What are the flight schedules to Paris tomorrow?'', voice assistants combine a brief voice-based response with a visual layout of multiple possible routes to maximize information throughput. 

Working with mixed-modal interactions effectively requires more research into understanding the differences between voice- and text-based queries and responses, especially for exploratory search.
In terms of query length, we do know that people are issuing longer voice-based queries that are more similar to natural language as compared to text-based searches~\cite{guy2018characteristics}, while earlier work reports that voice-based queries were shorter potentially due to perceptions of lower speech recognition accuracy~\cite{schalkwyk2010your}.
However, more analysis is needed to understand whether people use voice- and text-based search for different tasks (in particular information look up v.s. exploratory search), and whether this might change over time over perceived quality.

From the response side, we know relatively little about the effectiveness of different modes of responses in supporting exploratory search.
It is also an open question in terms of user preference --- do some users prefer voice-based while others prefer visual-based for the same tasks?
Future work is necessary for more rigorous comparisons of voice- and text-based search and responses --- for example, by constructing a voice- and text-search dataset for the same exploratory search tasks and evaluating the difference in efficiency for these tasks.
Answering these questions could help us better design and mix different modes of interactions for the best results.

\subsection{Designing for Diverse Population}
Voice assistants are used globally.
Hence, it is important to consider the diverse populations for whom we are designing.
These user groups can be eclectic in locations, devices, demographics, socioeconomic status, and both digital and language literacies.

Voice-based exploratory search may be particularly meaningful for certain user groups.
Prior research has suggested that literacy, gender, and language familiarity all affect how people interact with voice assistants~\cite{bhalla2018exploratory}.
In one study, low-income women who were semiliterate preferred to use voice searching, while men still preferred text-based searching~\cite{bhalla2018exploratory}.
Meanwhile, low-income population may use types of devices different that are not flagship devices and have limited support for visual elements.
Finally, prior literature noted that low-income users have an ``aspirational association with English''~\cite{karusala2018only,bhalla2018exploratory}.
Such aspirational association leads to users searching in English rather than in the vernacular, which may limit their experience.
It is important to bear these differences and constraints in mind when designing voice assistants that support exploratory search.

\subsection{Meeting Expectations and Gaining Trust}
Finally, user expectations and trust is the biggest hurdle.
We briefly mentioned above that user expectations for voice assistants have grown over the years with the advancement of speech technology, resulting in longer and more natural sounding queries.
In fact, the expectation has grown so high~\cite{chopra2017phone,luger2016like,tabassum2019investigating} that current systems cannot keep up, resulting in ``gulfs of execution and evaluation''~\cite{luger2016like}.

One hypothesis is that the anthropomorphism of voice assistants creates higher expectations and trust from users compared to text search, but that trust is more brittle and users are less forgiving towards assistants when they make mistakes.
To anthropomorphize is to ``ascribe human-like features and characteristics to an otherwise non-human object''~\cite{marakas2000theoretical,kontogiorgos2019effects,leong2019robot}.
Voice assistants, through anthropomorphism, may result in higher expectations from the users for exploratory search.
Users may expect the voice assistant to handle and resolve (through natural dialogs) ambiguity better than text-based search.
Future work is needed to test this hypothesis, especially around how technological literacy might play a role~\cite{luger2016like}.
On the other hand, such high expectations might make people less forgiving when the system makes a mistake --- users may lose trust quickly if the assistant fails to meet those expectations.
Even worse, users may never try similar tasks again once the trust is lost.
Therefore, it is important for voice assistants to reliably support exploratory search to maintain trust.

\section{Conclusion}

Many open questions still remain when it comes to designing a well-versed voice assistant that supports exploratory search more naturally.
In this position paper, we laid out four challenges around designing such voice assistants: addressing situationally induced impairments; working with mixed-modal interactions; designing for diverse populations; and meeting users' expectations and gaining their trust.
By addressing these challenges, we can best bridge the ``gulfs of execution and evaluation''~\cite{luger2016like} and meet the high user expectations for exploratory search through voice assistants.

\balance{} 

\bibliographystyle{SIGCHI-Reference-Format}
\bibliography{main}


\begin{thebibliography}{00}


\ifx \showCODEN    \undefined \def \showCODEN     #1{\unskip}     \fi
\ifx \showDOI      \undefined \def \showDOI       #1{{\tt DOI:}\penalty0{#1}\ }
  \fi
\ifx \showISBNx    \undefined \def \showISBNx     #1{\unskip}     \fi
\ifx \showISBNxiii \undefined \def \showISBNxiii  #1{\unskip}     \fi
\ifx \showISSN     \undefined \def \showISSN      #1{\unskip}     \fi
\ifx \showLCCN     \undefined \def \showLCCN      #1{\unskip}     \fi
\ifx \shownote     \undefined \def \shownote      #1{#1}          \fi
\ifx \showarticletitle \undefined \def \showarticletitle #1{#1}   \fi
\ifx \showURL      \undefined \def \showURL       #1{#1}          \fi

\bibitem{ammari2019music}
{Tawfiq Ammari}, {Jofish Kaye}, {Janice~Y. Tsai}, {and} {Frank Bentley}. 2019.
\newblock \showarticletitle{Music, Search, and IoT: How People (Really) Use
  Voice Assistants}.
\newblock {\em ACM Trans. Comput.-Hum. Interact.\/} {26}, 3, Article Article 17
  (April 2019), 28 pages.
\newblock
\showISSN{1073-0516}
\showDOI{%
\url{http://dx.doi.org/10.1145/3311956}}


\bibitem{bhalla2018exploratory}
{Apoorva Bhalla}. 2018.
\newblock \showarticletitle{An Exploratory Study Understanding the Appropriated
  Use of Voice-Based Search and Assistants}. In {\em Proceedings of the 9th
  Indian Conference on Human Computer Interaction} {\em (IndiaHCI'18)}.
  Association for Computing Machinery, New York, NY, USA, 90--94.
\newblock
\showISBNx{9781450362146}
\showDOI{%
\url{http://dx.doi.org/10.1145/3297121.3297136}}


\bibitem{cho2019hey}
{Eugene Cho}. 2019.
\newblock \showarticletitle{Hey Google, Can I Ask You Something in Private?}.
  In {\em Proceedings of the 2019 CHI Conference on Human Factors in Computing
  Systems} {\em (CHI '19)}. Association for Computing Machinery, New York, NY,
  USA, Article Paper 258, 9 pages.
\newblock
\showISBNx{9781450359702}
\showDOI{%
\url{http://dx.doi.org/10.1145/3290605.3300488}}


\bibitem{chopra2017phone}
{Simran Chopra} {and} {Shruthi Chivukula}. 2017.
\newblock \showarticletitle{My Phone Assistant Should Know I Am an Indian:
  Influencing Factors for Adoption of Assistive Agents}. In {\em Proceedings of
  the 19th International Conference on Human-Computer Interaction with Mobile
  Devices and Services} {\em (MobileHCI '17)}. Association for Computing
  Machinery, New York, NY, USA, Article Article 94, 8 pages.
\newblock
\showISBNx{9781450350754}
\showDOI{%
\url{http://dx.doi.org/10.1145/3098279.3122137}}


\bibitem{guy2018characteristics}
{Ido Guy}. 2018.
\newblock \showarticletitle{The Characteristics of Voice Search: Comparing
  Spoken with Typed-in Mobile Web Search Queries}.
\newblock {\em ACM Trans. Inf. Syst.\/} {36}, 3, Article Article 30 (March
  2018), 28 pages.
\newblock
\showISSN{1046-8188}
\showDOI{%
\url{http://dx.doi.org/10.1145/3182163}}


\bibitem{karusala2018only}
{Naveena Karusala}, {Aditya Vishwanath}, {Aditya Vashistha}, {Sunita Kumar},
  {and} {Neha Kumar}. 2018.
\newblock \showarticletitle{``Only If You Use English You Will Get to More
  Things'': Using Smartphones to Navigate Multilingualism}. In {\em Proceedings
  of the 2018 CHI Conference on Human Factors in Computing Systems} {\em (CHI
  '18)}. Association for Computing Machinery, New York, NY, USA, Article Paper
  573, 14 pages.
\newblock
\showISBNx{9781450356206}
\showDOI{%
\url{http://dx.doi.org/10.1145/3173574.3174147}}


\bibitem{kontogiorgos2019effects}
{Dimosthenis Kontogiorgos}, {Andre Pereira}, {Olle Andersson}, {Marco
  Koivisto}, {Elena Gonzalez~Rabal}, {Ville Vartiainen}, {and} {Joakim
  Gustafson}. 2019.
\newblock \showarticletitle{The Effects of Anthropomorphism and Non-Verbal
  Social Behaviour in Virtual Assistants}. In {\em Proceedings of the 19th ACM
  International Conference on Intelligent Virtual Agents} {\em (IVA '19)}.
  Association for Computing Machinery, New York, NY, USA, 133--140.
\newblock
\showISBNx{9781450366724}
\showDOI{%
\url{http://dx.doi.org/10.1145/3308532.3329466}}


\bibitem{leong2019robot}
{Brenda Leong} {and} {Evan Selinger}. 2019.
\newblock \showarticletitle{Robot Eyes Wide Shut: Understanding Dishonest
  Anthropomorphism}. In {\em Proceedings of the Conference on Fairness,
  Accountability, and Transparency} {\em (FAT* '19)}. Association for Computing
  Machinery, New York, NY, USA, 299--308.
\newblock
\showISBNx{9781450361255}
\showDOI{%
\url{http://dx.doi.org/10.1145/3287560.3287591}}


\bibitem{luger2016like}
{Ewa Luger} {and} {Abigail Sellen}. 2016.
\newblock \showarticletitle{``Like Having a Really Bad PA'': The Gulf between
  User Expectation and Experience of Conversational Agents}. In {\em
  Proceedings of the 2016 CHI Conference on Human Factors in Computing Systems}
  {\em (CHI '16)}. Association for Computing Machinery, New York, NY, USA,
  5286--5297.
\newblock
\showISBNx{9781450333627}
\showDOI{%
\url{http://dx.doi.org/10.1145/2858036.2858288}}


\bibitem{marakas2000theoretical}
{GEORGE~M. MARAKAS}, {RICHARD~D. JOHNSON}, {and} {JONATHAN~W. PALMER}. 2000.
\newblock \showarticletitle{A Theoretical Model of Differential Social
  Attributions toward Computing Technology}.
\newblock {\em Int. J. Hum.-Comput. Stud.\/} {52}, 4 (April 2000), 719--750.
\newblock
\showISSN{1071-5819}
\showDOI{%
\url{http://dx.doi.org/10.1006/ijhc.1999.0348}}


\bibitem{marchionini2006exploratory}
{Gary Marchionini}. 2006.
\newblock \showarticletitle{Exploratory Search: From Finding to Understanding}.
\newblock {\em Commun. ACM\/} {49}, 4 (April 2006), 41--46.
\newblock
\showISSN{0001-0782}
\showDOI{%
\url{http://dx.doi.org/10.1145/1121949.1121979}}


\bibitem{schalkwyk2010your}
{Johan Schalkwyk}, {Doug Beeferman}, {Fran{\c{c}}oise Beaufays}, {Bill Byrne},
  {Ciprian Chelba}, {Mike Cohen}, {Maryam Kamvar}, {and} {Brian Strope}. 2010.
\newblock \showarticletitle{``Your word is my command'': Google search by
  voice: A case study}.
\newblock In {\em Advances in speech recognition}. Springer, 61--90.
\newblock


\bibitem{tabassum2019investigating}
{Madiha Tabassum}, {Tomasz Kosiundefinedski}, {Alisa Frik}, {Nathan Malkin},
  {Primal Wijesekera}, {Serge Egelman}, {and} {Heather~Richter Lipford}. 2019.
\newblock \showarticletitle{Investigating Users' Preferences and Expectations
  for Always-Listening Voice Assistants}.
\newblock {\em Proc. ACM Interact. Mob. Wearable Ubiquitous Technol.\/} {3}, 4,
  Article Article 153 (Dec. 2019), 23 pages.
\newblock
\showDOI{%
\url{http://dx.doi.org/10.1145/3369807}}


\bibitem{vtyurina2019towards}
{Alexandra Vtyurina}. 2019.
\newblock \showarticletitle{Towards Non-Visual Web Search}. In {\em Proceedings
  of the 2019 Conference on Human Information Interaction and Retrieval} {\em
  (CHIIR '19)}. Association for Computing Machinery, New York, NY, USA,
  429--432.
\newblock
\showISBNx{9781450360258}
\showDOI{%
\url{http://dx.doi.org/10.1145/3295750.3298976}}


\end{thebibliography}

\end{document}